\documentclass{aa}  

\usepackage{graphicx}
\usepackage{txfonts}
\usepackage{lipsum}
\usepackage{subcaption}         
\usepackage{lscape}             
\usepackage{placeins}           
                                
\begin{document}
   \title{Binary Evolution Can Mimic the Pair-Instability Mass Gap in Black Hole Mergers}

%

\author{
Aleksandra Olejak\inst{1, 2}
\thanks{Corresponding author: aolejak@mpa-garching.mpg.de}
\and Jakob Stegmann \inst{1} 
}

\institute{
Max Planck Institute for Astrophysics,
Karl-Schwarzschild-Straße~1,
85748 Garching b.~München, Germany
 \and
 Nicolaus Copernicus Astronomical Center, Polish Academy of Sciences, ul. Bartycka 18, 00-716 Warsaw, Poland
}

   \date{Received September 30, 20XX}

 
  \abstract
{The recent O4 catalogs from the LIGO–Virgo–KAGRA collaboration, which significantly increased the number of gravitational-wave (GW) detections, reveal features with potentially important astrophysical implications. One notable example is a hint of the so-called pair-instability mass gap. In particular, the observed decline in the number of black holes (BHs) with mass above $\sim 45\,M_\odot$, together with indications of possibly higher spins for BHs above this threshold, has been interpreted by Antonini et al. and Tong et al. as evidence for pair-instability supernovae.}
{In this work, we investigate whether mass transfer in binary systems can produce features in the BH component-mass distributions that mimic the inferred pair-instability limit.}
{We use both the population synthesis code {\tt StarTrack} and a simple semi-analytical framework to highlight the impact of mass transfer efficiency on the BH masses.}
{We find that mass-ratio reversal during the first Roche-lobe overflow, followed by a highly non-conservative second mass-transfer phase, naturally limits the mass of the first-born BH and produces an apparent cutoff that mimics the lower edge of a mass gap. While the upper mass limit for the more massive BH is increased through accretion during the first mass-transfer phase and is ultimately set by the pair-instability limit, the less massive BH is limited to the stripped primary mass. As a result, the fraction of systems in which the less massive BH exceeds $\sim 45\,M_{\odot}$ is negligible, below $10^{-4}$ in our default models.}
{While pair instability remains a possible interpretation of current GW data, similar features can naturally arise from binary evolution without invoking a low pair-instability cutoff. The emerging evidence for predominantly positive effective spins extending into the putative upper mass gap may further support a contribution from isolated binary evolution to the high-mass tail.}


   \maketitle
\nolinenumbers

\section{Introduction}

Several hundred binary black hole (BBH) mergers have now been reported by the LIGO–Virgo–KAGRA (LVK) Collaboration, with the latest GWTC-5.0 catalog extending observations through the O4b phase of the fourth observing run \citep{LVK2025catalog,LVKcatlog2026}.
Analyses of the properties of these systems, such as mass and spin distributions of compact objects, reveal encoded information about their formation channels \citep{LVK2025populations, LVKproprties2026}. The ultimate goal is to distinguish the relative contributions of various formation environments of detected GW systems which might be isolated field evolution of binaries \citep[e.g.,][]{Belczynski2020,Bavera2020,Marchant2021, Olejak2021a, Briel2021, vanSon2021, Xu2025, Klencki2026} or higher multiples systems \citep[e.g.,][]{Antonini2016, Dorozsmai2024, Stegmann2025c, Stegmann2025} as well as globular clusters or galactic nuclei \citep[e.g.,][]{Antonini2016, Askar2017, Rodriguez2018b, Mapelli2021a, Giant2026}. Some recent analyses suggest the current GW population could consist of several subpopulations characterized by distinct features \citep{Li2025, Tong2025b, Afroz2025, Ray2026,Llobera-Querol2026, Cheng2026, Galaudage2026, Rinaldi2026}, which could point towards the contribution of various formation channels. 

It has been predicted for decades that the mass distribution of stellar-origin BHs should exhibit a gap caused by so-called pair-instability supernovae (PSN), originally expected to lie between around $50$ and $130\,M_{\odot}$ \citep{Fraley1968, Heger2002, Woosley2007, Belczynski2016}. The exact boundaries of this mass gap, however, remain highly uncertain, as they depend sensitively on factors such as the poorly constrained ${}^{12}$C($\alpha,\gamma$)${}^{16}$O reaction rate. As a result, the lower edge of the gap may vary from $\sim 35$ to even $\sim 90\,M_{\odot}$, while the upper edge may range from $\sim 130$ to $\sim 180\,M_{\odot}$ \citep{Farmer2020, Costa2021, Farag2022}. Nevertheless, this mechanism is expected to limit the formation of massive BHs formed through single stellar evolution above the uncertain threshold. Therefore, if PSN operates, the detected population of merging BBHs should exhibit a clearly distinguishable imprint of this mechanism in their mass distribution \citep{Belczynski2016}.

Among GW detections, there are mergers with BH components confidently lying within the expected PSN mass gap, such as {\tt GW190521} \citep{2020PhRvL.125j1102A} and the more recent cases of {\tt GW231028} and {\tt GW231005} \citep{LVK2025O4aCatalog}. However, those detections do not necessarily yet challenge PSN theory, as they could have a hierarchical origin \citep{Mapelli2021a, Gerosa2021, Kimball2021}. In particular, they might be the products of previous BBH mergers that subsequently paired in dense stellar environments, such as globular clusters, with another BH. BHs that formed through such successive mergers, the so-called second-generation BBH mergers, are expected to be characterized by high spins, with individual spin magnitudes peaking around $\sim 0.7$ \citep{Gerosa2017, Gerosa2021}. Unfortunately, individual BH spin magnitudes and spin-orbit misalignments inferred from GW data carry large uncertainties \citep{Vitale2014, Wolfe2026}, making it difficult to confidently confirm or rule out a second‑generation origin based on their values.

Recent analyses of GW data by \citet{Antonini2026} find support for a transition in the spin distribution of BHs associated with a high mass threshold of $\sim 40\,M_{\odot}$. In particular, the effective spins\footnote{Effective spin 
$
\chi_{\rm eff} = \frac{\chi_1 \cos\theta_1 + q,\chi_2 \cos\theta_2}{1+q},
$
where $\chi_{1(2)}$ and $\theta_{1(2)}$ are the component spin magnitudes and tilt angles, respectively, and $0 < q \le 1$ is the binary mass ratio~\citep{Damour2001}.} of the massive BHs within the apparent PSN mass‑gap region appear systematically higher than those of the lower‑mass BH subpopulation. Moreover, \citet{Tong2026} shows that when the mass distributions of the more massive and less massive BH components are reconstructed separately, the distribution of the lower-mass component reveals a gap starting at $\sim 45\,M_{\odot}$. Such features may indicate that BHs with masses above this transition are predominantly second-generation merger remnants. Taken together, these results can be interpreted as evidence for the PSN mass gap, with the lower edge estimated at $45.3^{+6.5}_{-4.8}\,M_{\odot}$ \citep{Antonini2026} and $45^{+5}_{-4}\,M_{\odot}$ \citep{Tong2026} with 90\% credibility. If this PSN edge is real, it provides constraints on the uncertain reaction rates and particle physics in very massive stellar cores \citep{Farmer2020, Costa2021, Farag2022, Fiorillo2026}.

While the interpretation of GW data with a PSN mass gap at $\sim 45\,M_{\odot}$ is plausible, we explore and highlight an alternative explanation. In particular, we demonstrate that binary evolution involving stable mass transfer alone can produce a sharp cutoff in the lower-mass BH component below $\sim 45\,M_{\odot}$, even if the true PSN limit lies much higher. Finally, we discuss how detections of the most massive BBH mergers could help distinguish our scenario from the proposed interpretation of a PSN mass gap at $\sim 45\,M_{\odot}$ \citep{Mould2022, Antonini2025, Antonini2026, Tong2026}.


\section{Evolutionary scenario mimicking PSN mass gap}

   \begin{figure}[ht!]
   \centering
   \includegraphics[width=6cm]{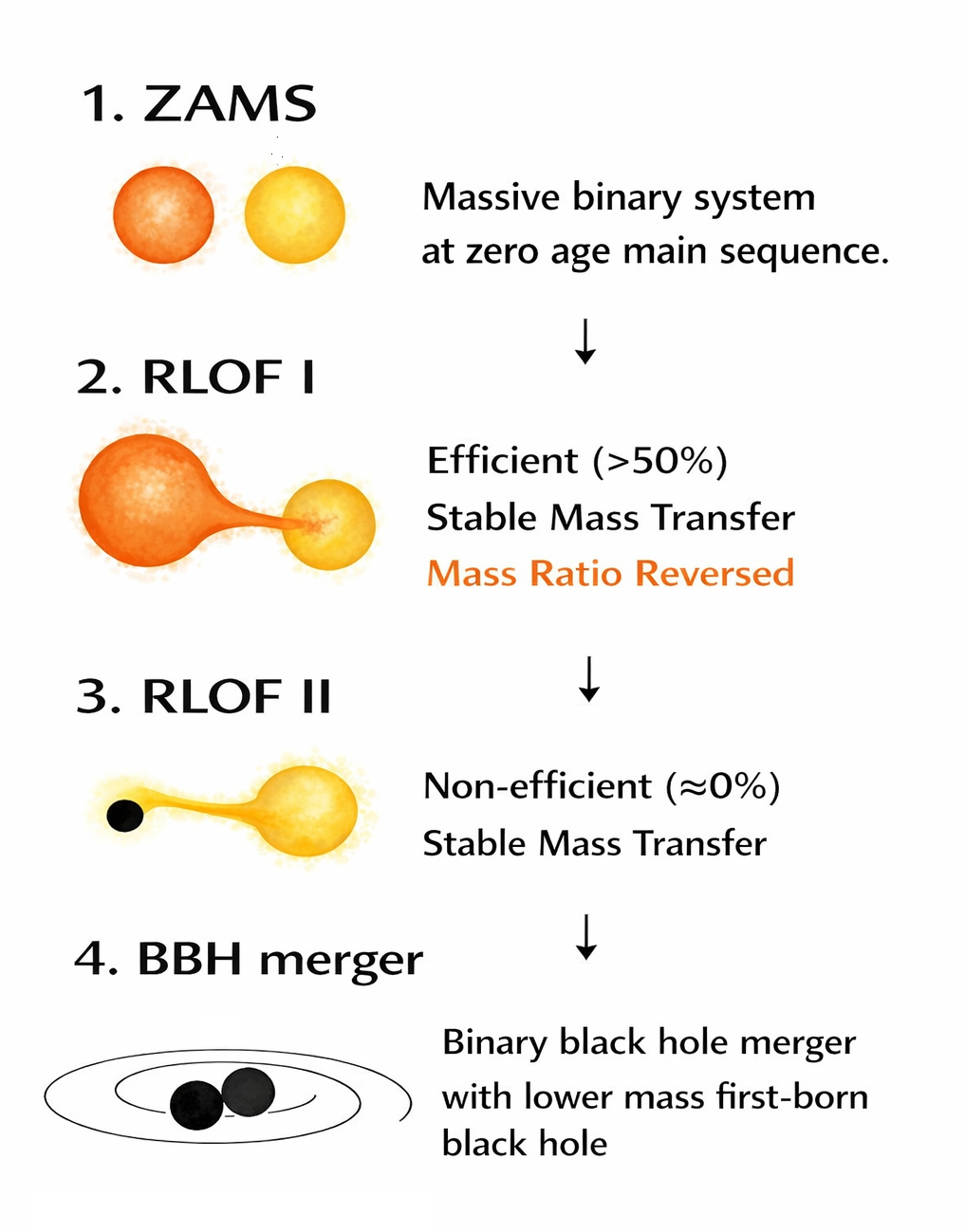}
      \caption{An evolutionary pathway for BBH mergers in which mass-transfer physics leads to a lower-mass first-born BH and a cutoff that mimics the PSN mass gap. 
      }
         \label{fig_evolution}
   \end{figure}

Figure \ref{fig_evolution} illustrates a schematic formation pathway for merging BBH systems that produces a cutoff in the less-massive BH mass distribution, thereby mimicking the lower edge of a PSN mass gap. The initially more massive star expands first and initiates mass transfer. During this first Roche-lobe overflow, the accretor gains a significant fraction of the transferred mass, leading to a mass-ratio reversal and orbital widening. The initially more massive star loses a substantial amount of mass through mass transfer and stellar winds, becomes stripped of its envelope, and eventually forms the first-born, less-massive BH. Subsequently, the secondary star, originally the less massive component but now significantly more massive following efficient accretion, expands and fills its Roche lobe. It transfers mass onto the first-born BH; however, because accretion onto the BH is limited to roughly the Eddington rate, this phase is highly non-conservative and contributes negligibly to the BH's mass growth. The system can therefore naturally end up with an unequal mass ratio, with the first-born BH being the less massive component. This evolutionary scenario has been presented and discussed in more detail in previous studies \citep[e.g.,][]{Olejak2021b, Olejak2024}.

\subsection{Synthetic population of binary black hole mergers}

We use the merging BBH population generated with the \texttt{StarTrack} population synthesis code \citep{Belczynski2008, Belczynski2020}. We incorporate the extended PSN limit following \cite{Belczynski2020PSN}, in which stars undergo PSN disruption if the final mass of their helium core lies in the range $90\,M_\odot \lesssim M_{\rm He} \lesssim 175\,M_\odot$. This model allows the formation of BHs with masses up to $\sim 90\,M_\odot$, consistent with the possible upper limit on the lower PSN edge \citep{Farmer2020, Costa2021}. We use revised mass-transfer stability criteria based on \cite{Pavlovskii2017}, which lead to the formation of BBH mergers predominantly through the stable mass-transfer channel \citep{Olejak2021a}. For the first mass-transfer phase, we consider two fixed mass-transfer efficiencies, $\beta=0.1$ and $0.5$. The remaining non-accreted mass is assumed to be lost with the specific orbital angular momentum of the accretor. The second mass-transfer phase onto the BH is highly non-conservative, with accretion limited to approximately the Eddington rate and modeled using the analytic approximations of \cite{King2001}.  Core-collapse supernova remnants are determined following the prescriptions of \citep{Fryer2022, Olejak2022}, adopting a mixing parameter of $f_{\rm mix}=4.0$ (rapid-type explosion model) and a critical pre-collapse core-mass threshold of $M_{\rm crit}=4.75\,M_\odot$ for BH formation.

\begin{figure}[ht!]
   \centering
   \includegraphics[width=\hsize]{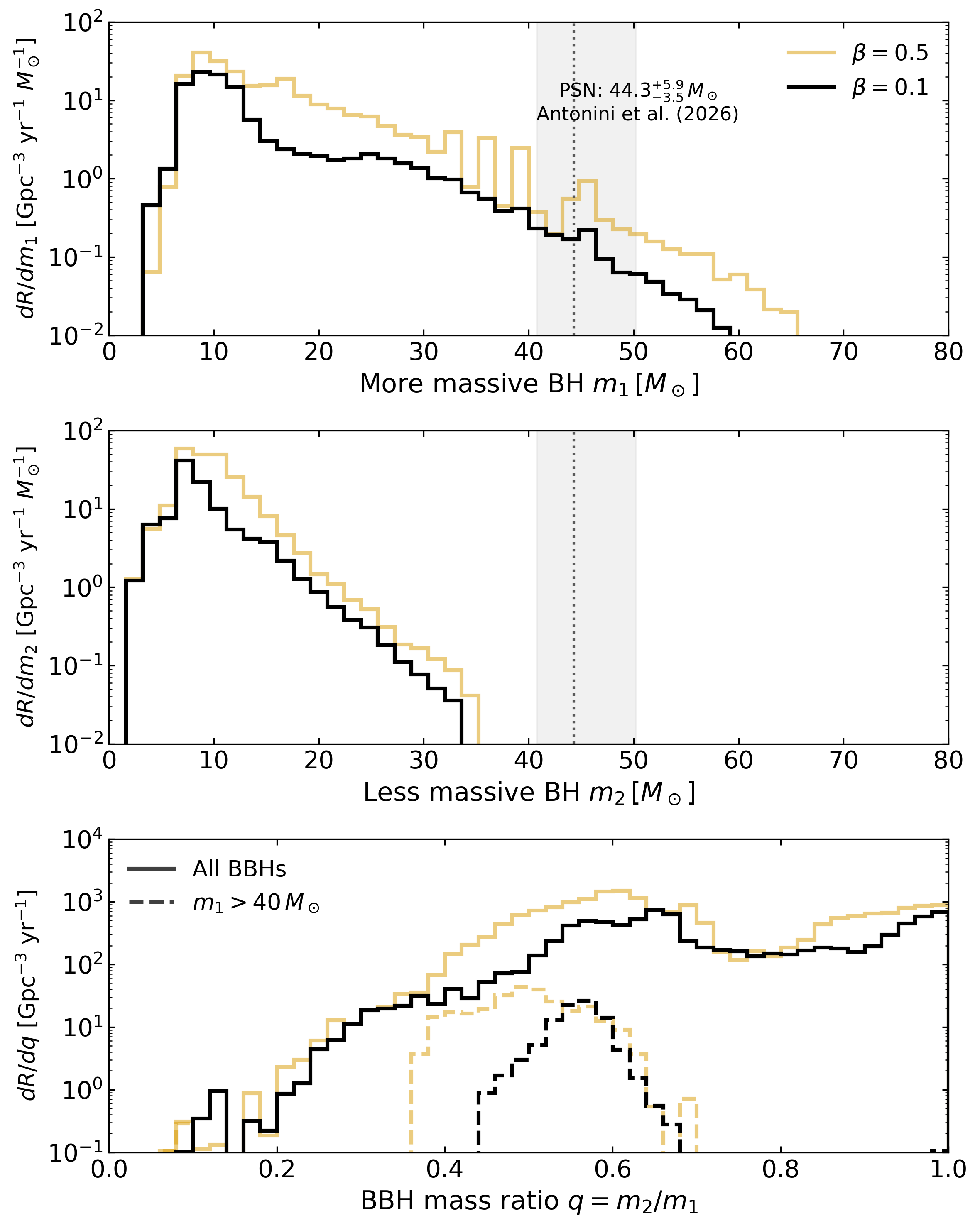}
    \caption{Intrinsic mass distributions of BBHs merging at redshift $z<0.3$. The upper and middle panels show the distributions of the more massive $m_1$ and less massive $m_2$ BH components, respectively, while the lower panel shows the binary mass-ratio distribution $q=m_2/m_1$. Black curves correspond to the model with low mass-transfer efficiency $\beta=0.1$, while gold curves show the alternative model with $\beta=0.5$. Solid lines represent the full BBH population, whereas dashed lines in the lower panel indicate the mass ratio distribution of massive systems with $m_1>40\, M_\odot$. The shaded region marks the inferred PSN mass limit, $(44.3_{-3.5}^{+5.9}\,M_{\odot}$ from \citet{Antonini2026}.
}
         \label{fig_mass_distribution}
   \end{figure}

\begin{table}
\centering
\caption{Fractions (in \%) of merging BBHs with each BH component exceeding selected mass thresholds of $45$, $50$, $55$, and $60\,M_{\odot}$, for two mass-transfer-efficiency models, $\beta=0.1$ and $\beta=0.5$, assuming a high PSN limit of $\sim 90\,M_{\odot}$.
}
\label{tab:mass_gap_fractions}
\begin{tabular}{c c c c c}
\hline\hline
Threshold & 
\multicolumn{2}{c}{Intrinsic $\beta=0.1$ } &
\multicolumn{2}{c}{Intrinsic $\beta = 0.5$ } \\
$M_{\rm thr}$ [$M_\odot$] & 
$m_1 > M_{\rm thr}$ & $m_2 > M_{\rm thr}$ &
$m_1 > M_{\rm thr}$ & $m_2 > M_{\rm thr}$ \\
\hline
45  &     0.458\%   &    \textbf{0.004\%}  &     0.820\%    &   \textbf{0.001\%} \\
50  &     0.168\%   &    \textbf{0.002\%}  &     0.358\%    &   \textbf{0.001\%} \\
55  &     0.056\%   &    \textbf{0.001\%}  &     0.170\%    &   \textbf{0.001\%} \\
60  &     0.004\%   &    \textbf{0.000\%}  &     0.054\%    &   \textbf{0.000\%} \\
\hline
\end{tabular}
\end{table}

As shown in Figure \ref{fig_mass_distribution}, the distribution of the more massive BH component in merging BBH systems extends to higher masses than that of the less massive BH, with its upper limit ultimately set by the PSN prescription adopted in the model, at $\sim 90\,M_{\odot}$. The maximum mass of the more massive BH also depends on the mass-transfer efficiency. For $\beta=0.5$ (gold line), the distribution extends to higher masses than for $\beta=0.1$ (black line), as the secondary accretes more mass during the first mass-transfer phase and can subsequently form a more massive BH. In contrast, the less massive BH component, whose progenitor is stripped during the first mass-transfer phase, is restricted to substantially lower masses. Systems with less massive BHs above $40\,M_{\odot}$ are very rare in both models; note the logarithmic scale. Despite adopting a much higher PSN limit ($\sim 90\,M_{\odot}$), these results qualitatively reproduce the trends in the BH component-mass distributions shown in Fig.~1 of \cite{Tong2026}, which have been interpreted as evidence for a PSN mass gap at $\sim 45 M_{\odot}$. Our BBH population models, particularly the model with $\beta=0.1$, also exhibit a peak near $10\,M_{\odot}$ in both BH component-mass distributions, in agreement with the LVK population analysis \citep{LVK2025populations}.

Moreover, in the mass-ratio distributions, we show separately the full population of merging BBHs and the subset of massive systems with $m_1>40\,M_{\odot}$, indicated by dashed lines. This shows that the mass-ratio-reversal scenario with a high PSN limit also naturally produces a transition toward more unequal mass ratios at higher primary BH masses, qualitatively consistent with recent GW population inferences \citep{Ray2026,Cheng2026}.

Table~\ref{tab:mass_gap_fractions} lists the fractions of merging BBHs in which the more massive or less massive BH exceeds selected mass thresholds of $45$, $50$, $55$, and $60\,M_{\odot}$, for both mass-transfer-efficiency models adopting a high PSN limit. Notably, the intrinsic fraction of mergers with the less massive BH exceeding $45\,M_{\odot}$ is below $0.01\%$ in our models, despite the substantially higher PSN limit. These trends are also reproduced by the simple semi-analytical models presented in Section~\ref{sec: Semi}, demonstrating that they arise from the basic physics of binary mass transfer rather than from specific details of the population-synthesis calculations.

\subsection{Semi-analytical model} \label{sec: Semi}
Here we construct toy semi-analytical models to further illustrate how the scenario described above operates in simple terms. 

We generate $N = 2 \times 10^5$ zero-age main-sequence binaries. The primary masses $m_1$ are drawn from a power-law initial mass function with slope $\alpha = 2.3$ in the range $20\,M_\odot \leq m_1 \leq 150\,M_\odot$ \citep{Kroupa1993}. The mass ratio $q = m_2/m_1$ is sampled from a uniform distribution between 0.1 and 1.0, and the secondary mass is set as $m_2 = q\,m_1$. We retain only systems in which both components exceed $20\,M_\odot$, corresponding to potential BBH progenitors.

We assume a fixed fraction of the stellar mass contained in the helium core, $f_{\mathrm{core}}$, equal to $0.35$ in our default model and $0.45$ in an alternative model. Each binary is evolved through two phases of mass transfer. During the first episode, the primary transfers a fraction $f_{\mathrm{MT}}$ of its envelope to the secondary. We test three cases: $f_{\mathrm{MT}} = 0.5,\ 0.7,\ 0.9$. The post-transfer stellar masses are
\begin{equation}
m_1' = f_{\mathrm{core}}\, m_1 , \quad
m_2' = m_2 + f_{\mathrm{MT}} (m_1 - f_{\mathrm{core}}\, m_1).
\end{equation}

The second mass-transfer episode is assumed to be fully non-conservative due to the Eddington-limited accretion onto the BH. We therefore assume that the secondary loses its entire envelope without increasing the mass of the primary. The primary mass remains unchanged, while the secondary is reduced to its core mass following the first mass-transfer episode,
\begin{equation}
m_1'' = m_1' , \quad
m_2'' = f_{\mathrm{core}}\, m_2'.
\end{equation}

The final BH masses are assumed to be $90\%$ of the final stellar masses, accounting for possible mass loss, e.g., due to stellar winds and during core collapse. In practice, stellar winds may remove a much larger fraction of mass from the system \citep[see e.g.,][]{Belczynski2010}.

For each surviving system, we define the heavier and lighter BH masses as
\begin{equation}
m_{\mathrm{H}} = \max(m_{\mathrm{BH},1}, m_{\mathrm{BH},2}), \quad
m_{\mathrm{L}} = \min(m_{\mathrm{BH},1}, m_{\mathrm{BH},2}),
\end{equation}
and the mass ratio
\begin{equation}
q = \frac{m_{\mathrm{L}}}{m_{\mathrm{H}}}.
\end{equation}

We impose a very high limit on the PSN mass gap by excluding systems in which either BH mass falls within
\begin{equation}
80\,M_\odot < m_{\mathrm{BH}} < 180\,M_\odot .
\end{equation}

\begin{figure}[ht!]
\centering
\includegraphics[width=\hsize]{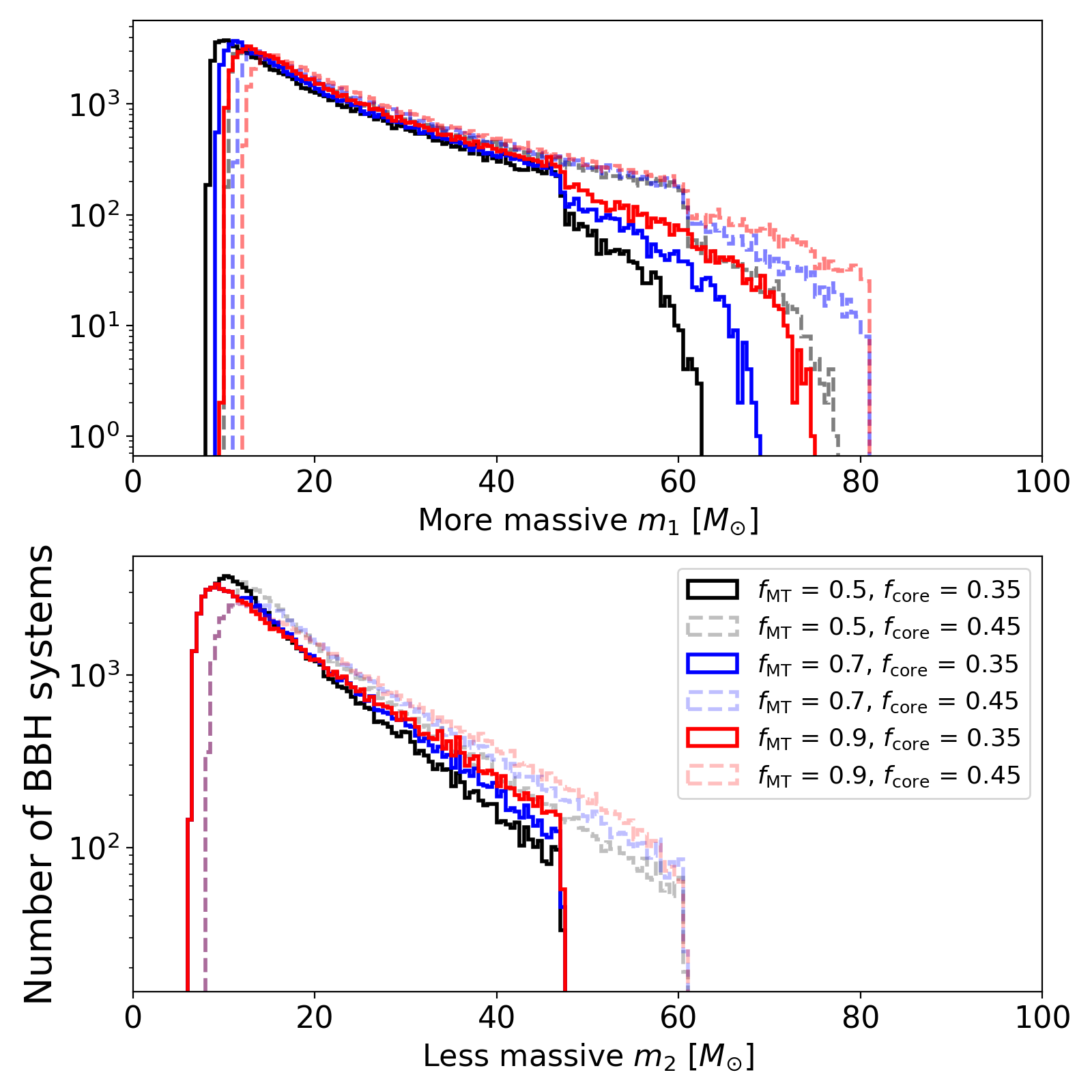}
\caption{Mass distribution of the more massive BH (upper panel), the less massive BH (bottom panel) of BBH mergers. Results are shown for semi-analytical models assuming three mass-transfer efficiencies: black — 50\%, blue — 70\%, and red — 90\%, and two helium-core mass fractions: 0.35 (solid lines) and 0.45 of the total stellar mass (dashed lines). The results demonstrate that the maximum mass of the more massive BH depends on the mass-transfer efficiency, the assumed core fraction, and the adopted high PSN limit. In contrast, the maximum mass of the less massive BH is determined only by the core fraction of the star.}
\label{fig_mass_distribution1}
\end{figure}

The described procedure is repeated for each value of the mass-transfer efficiency $f_{\mathrm{MT}}$, allowing us to assess its impact on the resulting BBH mass and mass-ratio distributions presented in Fig.~\ref{fig_mass_distribution1}.

We note that these simplified models do not incorporate several important physical processes that may affect the resulting distributions. For example, compared with the semi-analytical model, the population generated with \texttt{StarTrack} in Fig.~\ref{fig_mass_distribution} includes a small fraction of low-mass BHs ($<10\,M_{\odot}$) formed through successful or partially failed core-collapse supernovae according to the \cite{Fryer2022} prescription. Despite these simplifying assumptions, the semi-analytical models reproduce the key qualitative trend inferred from observed BBH populations: an apparent cutoff in the secondary-BH mass distribution \citep{LVK2025populations, Tong2026}.

We show that the mass of the more massive BH is primarily set by the amount of mass accreted during the first mass-transfer phase and, ultimately, by the assumed PSN limit. In contrast, the maximum mass of the less massive BH is set by the mass of the stripped core of the initially more massive star. For the two adopted core-mass fractions, $f_{\mathrm{core}}=0.35$ and $0.45$, this limit ranges from approximately $45$ to $60\,M_{\odot}$. Interestingly, some other analyses similarly suggest that the secondary-mass cutoff may occur at a higher mass, around $\sim 60\,M_{\odot}$ \citep{Ray2026b}.

\section{Discussion}

\subsection{Maximum BH masses from binary evolution}

As shown by our \texttt{StarTrack} and semi-analytical models, both the frequency of mass-ratio reversal and the maximum mass of the more massive BH component depend on the mass-transfer efficiency, i.e., the fraction of mass lost by the donor during the first mass-transfer phase that is accreted by the companion. Relatively efficient mass transfer is supported by constraints from several observed populations of post-mass-transfer binaries \citep{Vinciguerra2020,Romero-Shaw2023,Lechien2025,Sen2025}. However, efficient mass transfer is not commonly adopted as a default assumption in binary-evolution models of BBH formation. Many codes employ prescriptions with lower accretion efficiencies and substantial mass loss from the system during the first Roche-lobe overflow. In particular, accretion onto the companion is often limited by critical rotation \citep{deMink2007}, which can strongly suppress accretion and inhibit mass-ratio reversal, thereby favoring the formation of more equal-mass binaries, as typically found for systems undergoing Case A mass transfer \citep[e.g.,][]{Briel2026}.

Moreover, efficient mass transfer onto a non-compact accretor remains challenging to model with current state-of-the-art 1D stellar-evolution codes because of, for example, ``accretor swelling'' \citep{Schurmann2025}. This can lead to a contact phase whose subsequent evolution remains uncertain and may partly reflect limitations of the 1D treatment. Consequently, several studies of BBH formation using detailed 1D stellar-evolution calculations focus primarily on the second mass-transfer phase \citep{Gallegos-Garcia2021,Picco2024,Klencki2026}. As shown here, however, the first mass-transfer episode can play a crucial role in shaping both the BBH mass-ratio distribution and the apparent cutoff in the less-massive BH component. In reality, the mass-transfer efficiency is likely to vary significantly among binary systems, depending on properties such as the mass ratio, orbital parameters, and metallicity. Observations of some systems, including Gaia BH binaries and stripped Wolf--Rayet stars, may instead favor highly non-conservative mass-transfer phases \citep{Olejak2025,Xu2026}. A low mass-transfer efficiency can also help shift the BBH mass distribution toward the inferred $\sim10\,M_{\odot}$ peak, as demonstrated in this work and in other recent studies \citep{Chen2026,Maclean2026}.

Importantly, the maximum masses of both BH components also depend on the treatment of stellar winds, the adopted maximum zero-age main-sequence stellar mass, core overshooting, and other uncertain stellar-evolution parameters. Variations in these assumptions may further suppress the masses of both BH components below the $\sim45\,M_{\odot}$ threshold, rather than preferentially limiting only the less massive BH \citep{Iorio2023,Giacobbo2018,Bouffanais2021,Maclean2026}. This could make it even more challenging to distinguish between a PSN-induced mass gap and binary stripping as the physical origin of an apparent upper-mass cutoff in the BBH population.

\subsection{Black Hole Spins}

Another argument put forward by \cite{Antonini2026} and \cite{Tong2026} in favor of the PSN mass gap near $45\,M_{\odot}$ is a reported transition in the effective-spin distribution around a similar mass threshold toward higher values. Products of second-generation mergers are expected to produce a characteristic peak in the distribution of individual spin magnitudes around $\sim 0.7$ \citep{Gerosa2017}. Since individual spins are poorly constrained in GW observations, \cite{Antonini2026} and \cite{Tong2026} instead rely on the distribution of the effective-spin parameter $\chi_{\mathrm{eff}}$, which is generally measured more accurately but represents a degenerate combination of the component masses, spins, and their tilt angles relative to the orbital angular momentum. This approach comes with important caveats.

The robustness of $\chi_{\mathrm{eff}}$ measurements generally decreases toward higher total binary masses, as progressively fewer inspiral cycles remain in the detector band over which spin-dependent post-Newtonian phasing can accumulate \citep{Baird2013}. Nevertheless, $\chi_{\mathrm{eff}}$ also affects the late inspiral, merger, and properties of the remnant BH, and therefore measurable spin information can remain in merger--ringdown-dominated signals \citep{Haster2016}. At high masses, however, strong mass--ratio--spin degeneracies limit parameter constraints, while waveform systematics become increasingly important, as modelling inaccuracies in the merger--ringdown can produce biases comparable to or larger than statistical uncertainties even at moderate signal-to-noise ratios \citep{Baird2013,Purrer2013}. Consequently, although $\chi_{\mathrm{eff}}$ remains measurable for some high-mass systems, its constraints generally become broader and more model-dependent. This is illustrated in Figure ~5 of \cite{Antonini2026}, where several massive events exhibit broad uncertainty ranges in $\chi_{\mathrm{eff}}$.

Importantly, the most massive BBH events detected tend to exhibit predominantly positive effective spins, $\chi_{\mathrm{eff}}>0$ \citep{Wang2022,Li2025,Cheng2026,Rinaldi2026}, suggesting a preference for spins with a component aligned with the orbital angular momentum and, therefore, a potentially significant contribution from isolated binary evolution channels above the putative PSN mass gap \citep{Stegmann2025b}. By contrast, classical dynamical formation channels with isotropically oriented spins are expected to produce a more symmetric distribution of positive and negative $\chi_{\mathrm{eff}}$ values than currently inferred for the GW population \citep[see, e.g.,][]{Farr2017}. Moreover, recent population analyses find low-spin BHs extending to $\sim50$--$70\,M_{\odot}$, shifting the inferred upper cutoff of the low-spin population from $\sim 45\,M_{\odot}$ to $\sim65$--$70\,M_{\odot}$ \citep{Wang2026} and further challenging its direct association with the lower edge of the PSN mass gap.

The origin and magnitude of BH spins remain an active subject of debate (see the recent review by \citealt{Zdziarski2026}). One promising mechanism for producing rapidly spinning BHs through binary interactions is tidal spin-up of stripped helium stars in close post–mass-transfer binaries \citep{Belczynski2020, Bavera2020}. The mass-ratio reversal scenario presented here can naturally produce a more massive and more rapidly spinning BH \citep{Olejak2021b}. This process also gives rise to (anti)correlations between mass, mass ratio, and effective spin \citep{Broekgaarden2022, Olejak2024, BanerjeeOlejak2024}, which qualitatively resemble those inferred from GW observations \citep{LVKPop2023}. 
More recently \citet{Ma2026} revisited this mass-reversal BBH formation scenario using detailed stellar-evolution models of the second mass-transfer phase, finding that more massive donors preferentially undergo Case B mass transfer, leading to stronger tidal spin-up and consequently to more massive BBH mergers with higher effective spins.

\section{Summary}

In this work, we present an alternative to the PSN limit at $45\,M_{\odot}$ proposed, e.g., by \cite{Antonini2026} and \cite{Tong2026} to interpret current GW observations. In particular, we focus on explaining the high-mass tail of the BH component-mass distribution and the apparent cutoff in the less-massive BH distribution through binary interactions. We demonstrate that BBH mergers formed through a stable mass-transfer channel can reproduce the asymmetric mass distributions of the two BH components inferred by \cite{Tong2026}. This arises from the combination of (i) mass-ratio reversal during the first Roche-lobe overflow onto a non-compact accretor and (ii) highly non-conservative mass transfer onto the first-born BH during the second mass-transfer phase. These trends emerge consistently in both population-synthesis simulations with \texttt{StarTrack} and simple semi-analytical models. This formation scenario also leaves a characteristic imprint on the BBH mass-ratio distribution, producing a transition toward more unequal mass ratios at the high-mass end ($m_1\gtrsim40\,M_{\odot}$), qualitatively consistent with some recent GW population analyses.

Although spin measurements of massive BBH mergers remain uncertain, the relative fraction of negative effective spins could help distinguish between the PSN and binary-interaction interpretations. A roughly symmetric distribution of positive and negative $\chi_{\mathrm{eff}}$ within the putative mass gap would favor a substantial dynamical contribution, consistent with hierarchical mergers of PSN-limited first-generation BHs. Conversely, a preference for positive $\chi_{\mathrm{eff}}$ would support a contribution from isolated binary evolution extending to high BH masses. Current observations already provide tentative evidence in this direction, with the most massive BBH mergers preferentially exhibiting positive $\chi_{\mathrm{eff}}$ and population analyses suggesting an increasing contribution from spin-aligned systems at higher masses. Moreover, the presence of low-spin BHs within the putative mass gap would disfavor a predominantly second-generation origin, further weakening a direct association of the observed high-mass cutoff with the PSN mass gap. Future joint constraints on BH masses, mass ratios, and spins will provide a key test of these scenarios.

\begin{acknowledgements}
We thank Ilya Mandel, Ylva Götberg, Stephen Justham, Jakub Klencki, and Selma de Mink for useful discussions.
\end{acknowledgements}

\bibliographystyle{aa}
\bibliography{lib}

@ARTICLE{Belczynski2010,
       author = {{Belczynski}, Krzysztof and {Bulik}, Tomasz and {Fryer}, Chris L. and {Ruiter}, Ashley and {Valsecchi}, Francesca and {Vink}, Jorick S. and {Hurley}, Jarrod R.},
        title = "{On the Maximum Mass of Stellar Black Holes}",
      journal = {\apj},
         year = 2010,
        month = may,
       volume = {714},
       number = {2},
        pages = {1217-1226},
          doi = {10.1088/0004-637X/714/2/1217},
archivePrefix = {arXiv},
       eprint = {0904.2784},
 primaryClass = {astro-ph.SR},
       adsurl = {https://ui.adsabs.harvard.edu/abs/2010ApJ...714.1217B}
}

@ARTICLE{Belczynski2020,
       author = {{Belczynski}, K. and {Klencki}, J. and {Fields}, C.~E. and {Olejak}, A. and {Berti}, E. and {Meynet}, G. and {Fryer}, C.~L. and {Holz}, D.~E. and {O'Shaughnessy}, R. and {Brown}, D.~A. and {Bulik}, T. and {Leung}, S.~C. and {Nomoto}, K. and {Madau}, P. and {Hirschi}, R. and {Kaiser}, E. and {Jones}, S. and {Mondal}, S. and {Chruslinska}, M. and {Drozda}, P. and {Gerosa}, D. and {Doctor}, Z. and {Giersz}, M. and {Ekstrom}, S. and {Georgy}, C. and {Askar}, A. and {Baibhav}, V. and {Wysocki}, D. and {Natan}, T. and {Farr}, W.~M. and {Wiktorowicz}, G. and {Coleman Miller}, M. and {Farr}, B. and {Lasota}, J. -P.},
        title = "{Evolutionary roads leading to low effective spins, high black hole masses, and O1/O2 rates for LIGO/Virgo binary black holes}",
      journal = {\aap},
         year = 2020,
        month = apr,
       volume = {636},
          eid = {A104},
        pages = {A104},
          doi = {10.1051/0004-6361/201936528},
archivePrefix = {arXiv},
       eprint = {1706.07053},
 primaryClass = {astro-ph.HE},
       adsurl = {https://ui.adsabs.harvard.edu/abs/2020A&A...636A.104B}
}

@ARTICLE{Gerosa2021,
       author = {{Gerosa}, Davide and {Giacobbo}, Nicola and {Vecchio}, Alberto},
        title = "{High mass but low spin: an exclusion region to rule out hierarchical black-hole mergers as a mechanism to populate the pair-instability mass gap}",
      journal = {arXiv e-prints},
         year = 2021,
        month = apr,
          eid = {arXiv:2104.11247},
        pages = {arXiv:2104.11247},
archivePrefix = {arXiv},
       eprint = {2104.11247},
 primaryClass = {astro-ph.HE},
       adsurl = {https://ui.adsabs.harvard.edu/abs/2021arXiv210411247G}
}

@ARTICLE{Woosley2007,
       author = {{Woosley}, S.~E. and {Blinnikov}, S. and {Heger}, Alexander},
        title = "{Pulsational pair instability as an explanation for the most luminous supernovae}",
      journal = {\nat},
         year = 2007,
        month = nov,
       volume = {450},
       number = {7168},
        pages = {390-392},
          doi = {10.1038/nature06333},
archivePrefix = {arXiv},
       eprint = {0710.3314},
 primaryClass = {astro-ph},
       adsurl = {https://ui.adsabs.harvard.edu/abs/2007Natur.450..390W}
}

@ARTICLE{Kroupa1993,
   author = {{Kroupa}, P. and {Tout}, C.~A. and {Gilmore}, G.},
    title = "{The distribution of low-mass stars in the Galactic disc}",
  journal = {\mnras},
     year = 1993,
    month = jun,
   volume = 262,
    pages = {545-587},
      doi = {10.1093/mnras/262.3.545},
   adsurl = {https://ui.adsabs.harvard.edu/abs/1993MNRAS.262..545K}
}

@ARTICLE{Belczynski2008,
   author = {{Belczynski}, K. and {Kalogera}, V. and {Rasio}, F.~A. and {Taam}, R.~E. and 
	{Zezas}, A. and {Bulik}, T. and {Maccarone}, T.~J. and {Ivanova}, N.
	},
    title = "{Compact Object Modeling with the StarTrack Population Synthesis Code}",
  journal = {\apjs},
   eprint = {astro-ph/0511811},
     year = 2008,
    month = jan,
   volume = 174,
    pages = {223-260},
      doi = {10.1086/521026},
   adsurl = {https://ui.adsabs.harvard.edu/abs/2008ApJS..174..223B}
}

@ARTICLE{Pavlovskii2017,
       author = {{Pavlovskii}, K. and {Ivanova}, N. and {Belczynski}, K. and {Van}, K.~X.},
        title = "{Stability of mass transfer from massive giants: double black hole binary formation and ultraluminous X-ray sources}",
      journal = {\mnras},
         year = "2017",
        month = "Feb",
       volume = {465},
       number = {2},
        pages = {2092-2100},
          doi = {10.1093/mnras/stw2786},
archivePrefix = {arXiv},
       eprint = {1606.04921},
 primaryClass = {astro-ph.HE},
       adsurl = {https://ui.adsabs.harvard.edu/abs/2017MNRAS.465.2092P}
}

@ARTICLE{Bavera2020,
       author = {{Bavera}, Simone S. and {Fragos}, Tassos and {Qin}, Ying and
         {Zapartas}, Emmanouil and {Neijssel}, Coenraad J. and {Mandel}, Ilya and
         {Batta}, Aldo and {Gaebel}, Sebastian M. and {Kimball}, Chase and
         {Stevenson}, Simon},
        title = "{The origin of spin in binary black holes. Predicting the distributions of the main observables of Advanced LIGO}",
      journal = {\aap},
         year = 2020,
        month = mar,
       volume = {635},
          eid = {A97},
        pages = {A97},
          doi = {10.1051/0004-6361/201936204},
archivePrefix = {arXiv},
       eprint = {1906.12257},
 primaryClass = {astro-ph.HE},
       adsurl = {https://ui.adsabs.harvard.edu/abs/2020A&A...635A..97B}
}

@ARTICLE{Antonini2016,
   author = {{Antonini}, F. and {Rasio}, F.~A.},
    title = "{Merging Black Hole Binaries in Galactic Nuclei: Implications for Advanced-LIGO Detections}",
  journal = {\apj},
archivePrefix = "arXiv",
   eprint = {1606.04889},
 primaryClass = "astro-ph.HE",
     year = 2016,
    month = nov,
   volume = 831,
      eid = {187},
    pages = {187},
      doi = {10.3847/0004-637X/831/2/187},
   adsurl = {http://adsabs.harvard.edu/abs/2016ApJ...831..187A}
}

@ARTICLE{Askar2017,
   author = {{Askar}, A. and {Szkudlarek}, M. and {Gondek-Rosi{\'n}ska}, D. and
	{Giersz}, M. and {Bulik}, T.},
    title = "{MOCCA-SURVEY Database - I. Coalescing binary black holes originating from globular clusters}",
  journal = {\mnras},
archivePrefix = "arXiv",
   eprint = {1608.02520},
 primaryClass = "astro-ph.HE",
     year = 2017,
    month = jan,
   volume = 464,
    pages = {L36-L40},
      doi = {10.1093/mnrasl/slw177},
   adsurl = {http://adsabs.harvard.edu/abs/2017MNRAS.464L..36A}
}

@ARTICLE{Rodriguez2018b,
        author = {{Rodriguez}, Carl L. and {Amaro-Seoane}, Pau and
{Chatterjee}, Sourav and
          {Rasio}, Frederic A.},
         title = "{Post-Newtonian Dynamics in Dense Star Clusters:
Highly Eccentric, Highly Spinning, and Repeated Binary Black Hole Mergers}",
       journal = {\prl},
          year = "2018",
         month = "Apr",
        volume = {120},
        number = {15},
           eid = {151101},
         pages = {151101},
           doi = {10.1103/PhysRevLett.120.151101},
archivePrefix = {arXiv},
        eprint = {1712.04937},
  primaryClass = {astro-ph.HE},
        adsurl = {https://ui.adsabs.harvard.edu/abs/2018PhRvL.120o1101R}
}

@ARTICLE{Belczynski2020PSN,
       author = {{Belczynski}, Krzysztof},
        title = "{The Most Ordinary Formation of the Most Unusual Double Black Hole Merger}",
      journal = {\apjl},
         year = 2020,
        month = dec,
       volume = {905},
       number = {2},
          eid = {L15},
        pages = {L15},
          doi = {10.3847/2041-8213/abcbf1},
archivePrefix = {arXiv},
       eprint = {2009.13526},
 primaryClass = {astro-ph.HE},
       adsurl = {https://ui.adsabs.harvard.edu/abs/2020ApJ...905L..15B}
}

@ARTICLE{Vinciguerra2020,
       author = {{Vinciguerra}, Serena and {Neijssel}, Coenraad J. and {Vigna-G{\'o}mez}, Alejandro and {Mandel}, Ilya and {Podsiadlowski}, Philipp and {Maccarone}, Thomas J. and {Nicholl}, Matt and {Kingdon}, Samuel and {Perry}, Alice and {Salemi}, Francesco},
        title = "{Be X-ray binaries in the SMC as indicators of mass-transfer efficiency}",
      journal = {\mnras},
         year = 2020,
        month = nov,
       volume = {498},
       number = {4},
        pages = {4705-4720},
          doi = {10.1093/mnras/staa2177},
archivePrefix = {arXiv},
       eprint = {2003.00195},
 primaryClass = {astro-ph.HE},
       adsurl = {https://ui.adsabs.harvard.edu/abs/2020MNRAS.498.4705V}
}

@ARTICLE{Marchant2021,
       author = {{Marchant}, Pablo and {Pappas}, Kaliro{\"e} M.~W. and {Gallegos-Garcia}, Monica and {Berry}, Christopher P.~L. and {Taam}, Ronald E. and {Kalogera}, Vicky and {Podsiadlowski}, Philipp},
        title = "{The role of mass transfer and common envelope evolution in the formation of merging binary black holes}",
      journal = {\aap},
         year = 2021,
        month = jun,
       volume = {650},
          eid = {A107},
        pages = {A107},
          doi = {10.1051/0004-6361/202039992},
archivePrefix = {arXiv},
       eprint = {2103.09243},
 primaryClass = {astro-ph.SR},
       adsurl = {https://ui.adsabs.harvard.edu/abs/2021A&A...650A.107M}
}

@ARTICLE{King2001,
       author = {{King}, A.~R. and {Davies}, M.~B. and {Ward}, M.~J. and {Fabbiano}, G. and {Elvis}, M.},
        title = "{Ultraluminous X-Ray Sources in External Galaxies}",
      journal = {\apjl},
         year = 2001,
        month = may,
       volume = {552},
       number = {2},
        pages = {L109-L112},
          doi = {10.1086/320343},
archivePrefix = {arXiv},
       eprint = {astro-ph/0104333},
 primaryClass = {astro-ph},
       adsurl = {https://ui.adsabs.harvard.edu/abs/2001ApJ...552L.109K}
}

@ARTICLE{Olejak2021a,
       author = {{Olejak}, A. and {Belczynski}, K. and {Ivanova}, N.},
        title = "{Impact of common envelope development criteria on the formation of LIGO/Virgo sources}",
      journal = {\aap},
         year = 2021,
        month = jul,
       volume = {651},
          eid = {A100},
        pages = {A100},
          doi = {10.1051/0004-6361/202140520},
archivePrefix = {arXiv},
       eprint = {2102.05649},
 primaryClass = {astro-ph.HE},
       adsurl = {https://ui.adsabs.harvard.edu/abs/2021A&A...651A.100O}
}

@ARTICLE{Farmer2020,
       author = {{Farmer}, R. and {Renzo}, M. and {de Mink}, S.~E. and {Fishbach}, M. and {Justham}, S.},
        title = "{Constraints from Gravitational-wave Detections of Binary Black Hole Mergers on the $^{12}$C({\ensuremath{\alpha}}, {\ensuremath{\gamma}})$^{16}$O Rate}",
      journal = {\apjl},
         year = 2020,
        month = oct,
       volume = {902},
       number = {2},
          eid = {L36},
        pages = {L36},
          doi = {10.3847/2041-8213/abbadd},
archivePrefix = {arXiv},
       eprint = {2006.06678},
 primaryClass = {astro-ph.HE},
       adsurl = {https://ui.adsabs.harvard.edu/abs/2020ApJ...902L..36F}
}

@ARTICLE{vanSon2021,
       author = {{van Son}, L.~A.~C. and {de Mink}, S.~E. and {Callister}, T. and {Justham}, S. and {Renzo}, M. and {Wagg}, T. and {Broekgaarden}, F.~S. and {Kummer}, F. and {Pakmor}, R. and {Mandel}, I.},
        title = "{The Redshift Evolution of the Binary Black Hole Merger Rate: A Weighty Matter}",
      journal = {\apj},
         year = 2022,
        month = may,
       volume = {931},
       number = {1},
          eid = {17},
        pages = {17},
          doi = {10.3847/1538-4357/ac64a3},
archivePrefix = {arXiv},
       eprint = {2110.01634},
 primaryClass = {astro-ph.HE},
       adsurl = {https://ui.adsabs.harvard.edu/abs/2022ApJ...931...17V}
}

@ARTICLE{Briel2021,
       author = {{Briel}, M.~M. and {Eldridge}, J.~J. and {Stanway}, E.~R. and {Stevance}, H.~F. and {Chrimes}, A.~A.},
        title = "{Estimating Transient Rates from Cosmological Simulations and BPASS}",
      journal = {arXiv e-prints},
         year = 2021,
        month = nov,
          eid = {arXiv:2111.08124},
        pages = {arXiv:2111.08124},
archivePrefix = {arXiv},
       eprint = {2111.08124},
 primaryClass = {astro-ph.CO},
       adsurl = {https://ui.adsabs.harvard.edu/abs/2021arXiv211108124B}
}

@ARTICLE{Olejak2021b,
       author = {{Olejak}, A. and {Belczynski}, K.},
        title = "{The Implications of High Black Hole Spins for the Origin of Binary Black Hole Mergers}",
      journal = {\apjl},
         year = 2021,
        month = nov,
       volume = {921},
       number = {1},
          eid = {L2},
        pages = {L2},
          doi = {10.3847/2041-8213/ac2f48},
archivePrefix = {arXiv},
       eprint = {2109.06872},
 primaryClass = {astro-ph.HE},
       adsurl = {https://ui.adsabs.harvard.edu/abs/2021ApJ...921L...2O}
}

@ARTICLE{Fryer2022,
       author = {{Fryer}, Chris L. and {Olejak}, Aleksandra and {Belczynski}, Krzysztof},
        title = "{The Effect of Supernova Convection On Neutron Star and Black Hole Masses}",
      journal = {\apj},
         year = 2022,
        month = jun,
       volume = {931},
       number = {2},
          eid = {94},
        pages = {94},
          doi = {10.3847/1538-4357/ac6ac9},
archivePrefix = {arXiv},
       eprint = {2204.13025},
 primaryClass = {astro-ph.HE},
       adsurl = {https://ui.adsabs.harvard.edu/abs/2022ApJ...931...94F}
}

@ARTICLE{Olejak2022,
       author = {{Olejak}, Aleksandra and {Fryer}, Chris L. and {Belczynski}, Krzysztof and {Baibhav}, Vishal},
        title = "{The role of supernova convection for the lower mass gap in the isolated binary formation of gravitational wave sources}",
      journal = {\mnras},
         year = 2022,
        month = oct,
       volume = {516},
       number = {2},
        pages = {2252-2271},
          doi = {10.1093/mnras/stac2359},
archivePrefix = {arXiv},
       eprint = {2204.09061},
 primaryClass = {astro-ph.HE},
       adsurl = {https://ui.adsabs.harvard.edu/abs/2022MNRAS.516.2252O}
}

@ARTICLE{Dorozsmai2024,
       author = {{Dorozsmai}, Andris and {Toonen}, Silvia},
        title = "{Importance of stable mass transfer and stellar winds for the formation of gravitational wave sources}",
      journal = {\mnras},
         year = 2024,
        month = jun,
       volume = {530},
       number = {4},
        pages = {3706-3739},
          doi = {10.1093/mnras/stae152},
archivePrefix = {arXiv},
       eprint = {2207.08837},
 primaryClass = {astro-ph.SR},
       adsurl = {https://ui.adsabs.harvard.edu/abs/2024MNRAS.530.3706D}
}

@ARTICLE{Broekgaarden2022,
       author = {{Broekgaarden}, Floor S. and {Stevenson}, Simon and {Thrane}, Eric},
        title = "{Signatures of Mass Ratio Reversal in Gravitational Waves from Merging Binary Black Holes}",
      journal = {\apj},
         year = 2022,
        month = oct,
       volume = {938},
       number = {1},
          eid = {45},
        pages = {45},
          doi = {10.3847/1538-4357/ac8879},
archivePrefix = {arXiv},
       eprint = {2205.01693},
 primaryClass = {astro-ph.HE},
       adsurl = {https://ui.adsabs.harvard.edu/abs/2022ApJ...938...45B}
}

@ARTICLE{Gallegos-Garcia2021,
       author = {{Gallegos-Garcia}, Monica and {Berry}, Christopher P.~L. and {Marchant}, Pablo and {Kalogera}, Vicky},
        title = "{Binary Black Hole Formation with Detailed Modeling: Stable Mass Transfer Leads to Lower Merger Rates}",
      journal = {\apj},
         year = 2021,
        month = dec,
       volume = {922},
       number = {2},
          eid = {110},
        pages = {110},
          doi = {10.3847/1538-4357/ac2610},
archivePrefix = {arXiv},
       eprint = {2107.05702},
 primaryClass = {astro-ph.HE},
       adsurl = {https://ui.adsabs.harvard.edu/abs/2021ApJ...922..110G}
}

@ARTICLE{deMink2007,
       author = {{de Mink}, S.~E. and {Pols}, O.~R. and {Hilditch}, R.~W.},
        title = "{Efficiency of mass transfer in massive close binaries. Tests from double-lined eclipsing binaries in the SMC}",
      journal = {\aap},
         year = 2007,
        month = jun,
       volume = {467},
       number = {3},
        pages = {1181-1196},
          doi = {10.1051/0004-6361:20067007},
archivePrefix = {arXiv},
       eprint = {astro-ph/0703480},
 primaryClass = {astro-ph},
       adsurl = {https://ui.adsabs.harvard.edu/abs/2007A&A...467.1181D}
}

@ARTICLE{Picco2024,
       author = {{Picco}, A. and {Marchant}, P. and {Sana}, H. and {Nelemans}, G.},
        title = "{Forming merging double compact objects with stable mass transfer}",
      journal = {\aap},
         year = 2024,
        month = jan,
       volume = {681},
          eid = {A31},
        pages = {A31},
          doi = {10.1051/0004-6361/202347090},
archivePrefix = {arXiv},
       eprint = {2309.05736},
 primaryClass = {astro-ph.SR},
       adsurl = {https://ui.adsabs.harvard.edu/abs/2024A&A...681A..31P}
}

@ARTICLE{Farag2022,
       author = {{Farag}, Ebraheem and {Renzo}, Mathieu and {Farmer}, Robert and {Chidester}, Morgan T. and {Timmes}, F.~X.},
        title = "{Resolving the Peak of the Black Hole Mass Spectrum}",
      journal = {\apj},
         year = 2022,
        month = oct,
       volume = {937},
       number = {2},
          eid = {112},
        pages = {112},
          doi = {10.3847/1538-4357/ac8b83},
archivePrefix = {arXiv},
       eprint = {2208.09624},
 primaryClass = {astro-ph.HE},
       adsurl = {https://ui.adsabs.harvard.edu/abs/2022ApJ...937..112F}
}

@ARTICLE{Costa2021,
       author = {{Costa}, Guglielmo and {Bressan}, Alessandro and {Mapelli}, Michela and {Marigo}, Paola and {Iorio}, Giuliano and {Spera}, Mario},
        title = "{Formation of GW190521 from stellar evolution: the impact of the hydrogen-rich envelope, dredge-up, and $^{12}$C({\ensuremath{\alpha}}, {\ensuremath{\gamma}})$^{16}$O rate on the pair-instability black hole mass gap}",
      journal = {\mnras},
         year = 2021,
        month = mar,
       volume = {501},
       number = {3},
        pages = {4514-4533},
          doi = {10.1093/mnras/staa3916},
archivePrefix = {arXiv},
       eprint = {2010.02242},
 primaryClass = {astro-ph.SR},
       adsurl = {https://ui.adsabs.harvard.edu/abs/2021MNRAS.501.4514C}
}

@ARTICLE{Mapelli2021a,
       author = {{Mapelli}, Michela and {Dall'Amico}, Marco and {Bouffanais}, Yann and {Giacobbo}, Nicola and {Arca Sedda}, Manuel and {Artale}, M. Celeste and {Ballone}, Alessandro and {Di Carlo}, Ugo N. and {Iorio}, Giuliano and {Santoliquido}, Filippo and {Torniamenti}, Stefano},
        title = "{Hierarchical black hole mergers in young, globular and nuclear star clusters: the effect of metallicity, spin and cluster properties}",
      journal = {\mnras},
         year = 2021,
        month = jul,
       volume = {505},
       number = {1},
        pages = {339-358},
          doi = {10.1093/mnras/stab1334},
archivePrefix = {arXiv},
       eprint = {2103.05016},
 primaryClass = {astro-ph.HE},
       adsurl = {https://ui.adsabs.harvard.edu/abs/2021MNRAS.505..339M}
}

@ARTICLE{Antonini2026,
       author = {{Antonini}, Fabio and {Romero-Shaw}, Isobel M. and {Callister}, Thomas and {Dosopoulou}, Fani and {Chattopadhyay}, Debatri and {Ginat}, Yonadav Barry and {Gieles}, Mark and {Mapelli}, Michela},
        title = "{Gravitational-wave constraints on the pair-instability mass gap and nuclear burning in massive stars}",
      journal = {Nature Astronomy},
         year = 2026,
        month = jul,
       volume = {10},
        pages = {1049-1056},
          doi = {10.1038/s41550-026-02847-0},
archivePrefix = {arXiv},
       eprint = {2509.04637},
 primaryClass = {astro-ph.HE},
       adsurl = {https://ui.adsabs.harvard.edu/abs/2026NatAs..10.1049A}
}

@ARTICLE{Antonini2025,
       author = {{Antonini}, Fabio and {Callister}, Thomas and {Dosopoulou}, Fani and {Romero-Shaw}, Isobel M. and {Chattopadhyay}, Debatri},
        title = "{Inferring the pair-instability mass gap from gravitational wave data}",
      journal = {\prd},
         year = 2025,
        month = sep,
       volume = {112},
       number = {6},
          eid = {063040},
        pages = {063040},
          doi = {10.1103/nxnr-pdyx},
archivePrefix = {arXiv},
       eprint = {2506.09154},
 primaryClass = {astro-ph.HE},
       adsurl = {https://ui.adsabs.harvard.edu/abs/2025PhRvD.112f3040A}
}

@ARTICLE{Olejak2024,
       author = {{Olejak}, Aleksandra and {Klencki}, Jakub and {Xu}, Xiao-Tian and {Wang}, Chen and {Belczynski}, Krzysztof and {Lasota}, Jean-Pierre},
        title = "{Unequal-mass highly spinning binary black hole mergers in the stable mass transfer formation channel}",
      journal = {\aap},
         year = 2024,
        month = sep,
       volume = {689},
          eid = {A305},
        pages = {A305},
          doi = {10.1051/0004-6361/202450480},
archivePrefix = {arXiv},
       eprint = {2404.12426},
 primaryClass = {astro-ph.HE},
       adsurl = {https://ui.adsabs.harvard.edu/abs/2024A&A...689A.305O}
}

@ARTICLE{Sen2025,
       author = {{Sen}, Koushik and {Renzo}, Mathieu and {Jin}, Harim and {Langer}, Norbert and {Schootemeijer}, Abel and {Villase{\~n}or}, Jaime I. and {Mahy}, Laurent and {Grichener}, Aldana and {Shah}, Neev and {Wang}, Chen and {Xu}, Xiao-Tian},
        title = "{Interacting binaries on the Main Sequence as in-situ tracers of mass transfer efficiency and stability}",
      journal = {arXiv e-prints},
         year = 2025,
        month = nov,
          eid = {arXiv:2511.15347},
        pages = {arXiv:2511.15347},
          doi = {10.48550/arXiv.2511.15347},
archivePrefix = {arXiv},
       eprint = {2511.15347},
 primaryClass = {astro-ph.SR},
       adsurl = {https://ui.adsabs.harvard.edu/abs/2025arXiv251115347S}
}

@ARTICLE{Lechien2025,
       author = {{Lechien}, Thibault and {de Mink}, Selma E. and {Valli}, Ruggero and {Rubio}, Amanda C. and {van Son}, Lieke A.~C. and {Klement}, Robert and {Jin}, Harim and {Pols}, Onno},
        title = "{Binary Stars Take What They Get: Evidence for Efficient Mass Transfer from Stripped Stars with Rapidly Rotating Companions}",
      journal = {\apjl},
         year = 2025,
        month = sep,
       volume = {990},
       number = {2},
          eid = {L51},
        pages = {L51},
          doi = {10.3847/2041-8213/adfdd4},
archivePrefix = {arXiv},
       eprint = {2505.14780},
 primaryClass = {astro-ph.SR},
       adsurl = {https://ui.adsabs.harvard.edu/abs/2025ApJ...990L..51L}
}

@ARTICLE{LVK2025populations,
       author = {{The LIGO Scientific Collaboration} and {the Virgo Collaboration} and {the KAGRA Collaboration} and {Abac}, A.~G. and {Abouelfettouh}, I. and {Acernese}, F. and {Ackley}, K. and {Adamcewicz}, C. and {Adhicary}, S. and {Adhikari}, D. and {Adhikari}, N. and {Adhikari}, R.~X. and {Adkins}, V.~K. and {Afroz}, S. and {Agarwal}, D. and {Agathos}, M. and {Aghaei Abchouyeh}, M. and {Aguiar}, O.~D. and {Ahmadzadeh}, S. and {Aiello}, L. and {Ain}, A. and {Ajith}, P. and {Akutsu}, T. and {Albanesi}, S. and {Alfaidi}, R.~A. and {Al-Jodah}, A. and {All{\'e}n{\'e}}, C. and {Allocca}, A. and {Al-Shammari}, S. and {Altin}, P.~A. and {Alvarez-Lopez}, S. and {Amarasinghe}, O. and {Amato}, A. and {Amra}, C. and {Ananyeva}, A. and {Anderson}, S.~B. and {Anderson}, W.~G. and {Andia}, M. and {Ando}, M. and {Andrade}, T. and {Andr{\'e}s-Carcasona}, M. and {Andri{\'c}}, T. and {Anglin}, J. and {Ansoldi}, S. and {Antelis}, J.~M. and {Antier}, S. and {Aoumi}, M. and {Appavuravther}, E.~Z. and {Appert}, S. and {Apple}, S.~K. and {Arai}, K. and {Araya}, A. and {Araya}, M.~C. and {Arca Sedda}, M. and {Areeda}, J.~S. and {Argianas}, L. and {Aritomi}, N. and {Armato}, F. and {Armstrong}, S. and {Arnaud}, N. and {Arogeti}, M. and {Aronson}, S.~M. and {Arun}, K.~G. and {Ashton}, G. and {Aso}, Y. and {Assiduo}, M. and {Assis de Souza Melo}, S. and {Aston}, S.~M. and {Astone}, P. and {Attadio}, F. and {Aubin}, F. and {AultONeal}, K. and {Avallone}, G. and {Babak}, S. and {Badaracco}, F. and {Badger}, C. and {Bae}, S. and {Bagnasco}, S. and {Bagui}, E. and {Baiotti}, L. and {Bajpai}, R. and {Baka}, T. and {Baker}, T. and {Ball}, M. and {Ballardin}, G. and {Ballmer}, S.~W. and {Banagiri}, S. and {Banerjee}, B. and {Bankar}, D. and {Baptiste}, T.~M. and {Baral}, P. and {Barayoga}, J.~C. and {Barish}, B.~C. and {Barker}, D. and {Barman}, N. and {Barneo}, P. and {Barone}, F. and {Barr}, B. and {Barsotti}, L. and {Barsuglia}, M. and {Barta}, D. and {Bartoletti}, A.~M. and {Barton}, M.~A. and {Bartos}, I. and {Basak}, S. and {Basalaev}, A. and {Bassiri}, R. and {Basti}, A. and {Bates}, D.~E. and {Bawaj}, M. and {Baxi}, P. and {Bayley}, J.~C. and {Baylor}, A.~C. and {Baynard}, II, P.~A. and {Bazzan}, M. and {Bedakihale}, V.~M. and {Beirnaert}, F. and {Bejger}, M. and {Belardinelli}, D. and {Bell}, A.~S. and {Bellie}, D.~S. and {Bellizzi}, L. and {Beltran-Martinez}, D. and {Benoit}, W. and {Bentara}, I. and {Bentley}, J.~D. and {Ben Yaala}, M. and {Bera}, S. and {Bergamin}, F. and {Berger}, B.~K. and {Bernuzzi}, S. and {Beroiz}, M. and {Berry}, C.~P.~L. and {Bersanetti}, D. and {Bertolini}, A. and {Betzwieser}, J. and {Beveridge}, D. and {Bevilacqua}, G. and {Bevins}, N. and {Bhandare}, R. and {Bhatt}, R. and {Bhattacharjee}, D. and {Bhaumik}, S. and {Bhowmick}, S. and {Biancalana}, V. and {Bianchi}, A. and {Bilenko}, I.~A. and {Billingsley}, G. and {Binetti}, A. and {Bini}, S. and {Binu}, C. and {Birnholtz}, O. and {Biscoveanu}, S. and {Bisht}, A. and {Bitossi}, M. and {Bizouard}, M. -A. and {Blaber}, S. and {Blackburn}, J.~K. and {Blagg}, L.~A. and {Blair}, C.~D. and {Blair}, D.~G. and {Bobba}, F. and {Bode}, N. and {Boileau}, G. and {Boldrini}, M. and {Bolingbroke}, G.~N. and {Bolliand}, A. and {Bonavena}, L.~D. and {Bondarescu}, R. and {Bondu}, F. and {Bonilla}, E. and {Bonilla}, M.~S. and {Bonino}, A. and {Bonnand}, R. and {Booker}, P. and {Borchers}, A. and {Borhanian}, S. and {Boschi}, V. and {Bose}, S. and {Bossilkov}, V. and {Boudon}, A. and {Bozzi}, A. and {Bradaschia}, C. and {Brady}, P.~R. and {Branch}, A. and {Branchesi}, M. and {Braun}, I. and {Briant}, T. and {Brillet}, A. and {Brinkmann}, M. and {Brockill}, P. and {Brockmueller}, E. and {Brooks}, A.~F. and {Brown}, B.~C. and {Brown}, D.~D. and {Brozzetti}, M.~L. and {Brunett}, S. and {Bruno}, G. and {Bruntz}, R. and {Bryant}, J.},
        title = "{GWTC-4.0: Population Properties of Merging Compact Binaries}",
      journal = {arXiv e-prints},
         year = 2025,
        month = aug,
          eid = {arXiv:2508.18083},
        pages = {arXiv:2508.18083},
          doi = {10.48550/arXiv.2508.18083},
archivePrefix = {arXiv},
       eprint = {2508.18083},
 primaryClass = {astro-ph.HE},
       adsurl = {https://ui.adsabs.harvard.edu/abs/2025arXiv250818083T}
}

@ARTICLE{Fraley1968,
       author = {{Fraley}, Gary S.},
        title = "{Supernovae Explosions Induced by Pair-Production Instability}",
      journal = {\apss},
         year = 1968,
        month = aug,
       volume = {2},
       number = {1},
        pages = {96-114},
          doi = {10.1007/BF00651498},
       adsurl = {https://ui.adsabs.harvard.edu/abs/1968Ap&SS...2...96F}
}

@ARTICLE{Belczynski2016,
       author = {{Belczynski}, K. and {Heger}, A. and {Gladysz}, W. and {Ruiter}, A.~J. and {Woosley}, S. and {Wiktorowicz}, G. and {Chen}, H.-Y. and {Bulik}, T. and {O'Shaughnessy}, R. and {Holz}, D.~E. and {Fryer}, C.~L. and {Berti}, E.},
        title = "{The effect of pair-instability mass loss on black-hole mergers}",
      journal = {\aap},
         year = 2016,
        month = oct,
       volume = {594},
          eid = {A97},
        pages = {A97},
          doi = {10.1051/0004-6361/201628980},
archivePrefix = {arXiv},
       eprint = {1607.03116},
 primaryClass = {astro-ph.HE},
       adsurl = {https://ui.adsabs.harvard.edu/abs/2016A&A...594A..97B}
}

@ARTICLE{Heger2002,
       author = {{Heger}, A. and {Woosley}, S.~E.},
        title = "{The Nucleosynthetic Signature of Population III}",
      journal = {\apj},
         year = 2002,
        month = mar,
       volume = {567},
       number = {1},
        pages = {532-543},
          doi = {10.1086/338487},
archivePrefix = {arXiv},
       eprint = {astro-ph/0107037},
 primaryClass = {astro-ph},
       adsurl = {https://ui.adsabs.harvard.edu/abs/2002ApJ...567..532H}
}

@ARTICLE{Stegmann2025,
       author = {{Stegmann}, Jakob and {Antonini}, Fabio and {Olejak}, Aleksandra and {Biscoveanu}, Sylvia and {Raymond}, Vivien and {Rinaldi}, Stefano and {Flanagan}, Beth},
        title = "{In-plane Black-hole Spin Measurements Suggest Most Gravitational-wave Mergers Form in Triples}",
      journal = {arXiv e-prints},
         year = 2025,
        month = dec,
          eid = {arXiv:2512.15873},
        pages = {arXiv:2512.15873},
          doi = {10.48550/arXiv.2512.15873},
archivePrefix = {arXiv},
       eprint = {2512.15873},
 primaryClass = {astro-ph.HE},
       adsurl = {https://ui.adsabs.harvard.edu/abs/2025arXiv251215873S}
}

@ARTICLE{Ray2026,
       author = {{Ray}, Anarya and {Mukherjee}, Shirsha and {Zevin}, Michael and {Kalogera}, Vicky},
        title = "{On the Astrophysical Origin of Binary Black Hole Subpopulations: A Tale of Three Channels?}",
      journal = {arXiv e-prints},
         year = 2026,
        month = mar,
          eid = {arXiv:2603.17987},
        pages = {arXiv:2603.17987},
          doi = {10.48550/arXiv.2603.17987},
archivePrefix = {arXiv},
       eprint = {2603.17987},
 primaryClass = {astro-ph.HE},
       adsurl = {https://ui.adsabs.harvard.edu/abs/2026arXiv260317987R}
}

@ARTICLE{Tong2025b,
       author = {{Tong}, Hui and {Callister}, Thomas A. and {Fishbach}, Maya and {Thrane}, Eric and {Antonini}, Fabio and {Stevenson}, Simon and {Romero-Shaw}, Isobel M. and {Dosopoulou}, Fani},
        title = "{A subpopulation of low-mass, spinning black holes: signatures of dynamical assembly}",
      journal = {arXiv e-prints},
         year = 2025,
        month = nov,
          eid = {arXiv:2511.05316},
        pages = {arXiv:2511.05316},
          doi = {10.48550/arXiv.2511.05316},
archivePrefix = {arXiv},
       eprint = {2511.05316},
 primaryClass = {astro-ph.HE},
       adsurl = {https://ui.adsabs.harvard.edu/abs/2025arXiv251105316T}
}

@ARTICLE{Fiorillo2026,
       author = {{Fiorillo}, Damiano F.~G. and {Lucente}, Giuseppe and {Sakstein}, Jeremy and {Vitagliano}, Edoardo and {Cantiello}, Matteo},
        title = "{The Black Hole Mass Gap as a New Probe of Millicharged Particles}",
      journal = {arXiv e-prints},
         year = 2026,
        month = apr,
          eid = {arXiv:2604.02413},
        pages = {arXiv:2604.02413},
archivePrefix = {arXiv},
       eprint = {2604.02413},
 primaryClass = {hep-ph},
       adsurl = {https://ui.adsabs.harvard.edu/abs/2026arXiv260402413F}
}

\appendix
\nolinenumbers


\end{document}